\documentclass{jps-cp}
\usepackage{txfonts}
\usepackage{graphicx}
\usepackage{geometry}
\usepackage[detect-all]{siunitx}

\title{Four-channel System for Characterization of Josephson Parametric Amplifiers}

\author{Boris I. \textsc{Ivanov}$^{1}$, Jinmyeong \textsc{Kim}$^{2,1}$, Çağlar \textsc{Kutlu}$^{2,1}$, Arjan F. \textsc{van Loo}$^{3,4}$, Yasunobu~\textsc{Nakamura}$^{3,4}$, Sergey~V.~\textsc{Uchaikin}$^{1}$, Seonjeong~\textsc{Oh}$^{1}$, Violeta~\textsc{Gkika}$^{1}$, Andrei~\textsc{Matlashov}$^{1}$, Woohyun~\textsc{Chung}$^{1}$, and Yannis K. \textsc{Semertzidis}$^{1,2}$}

\inst{$^{1}$Center for Axion and Precision Physics Research, IBS, Daejeon 34051, Republic of Korea \\
$^{2}$Department of Physics, KAIST, Daejeon 34141, Republic of Korea \\
$^{3}$RIKEN Center for Quantum Computing (RQC), Wako, Saitama 351–0198, Japan \\
$^{4}$Department of Applied Physics, Graduate School of Engineering, The University of Tokyo, Bunkyo-ku, Tokyo 113-8656, Japan}

\email{b.ivanov@ibs.re.kr}

\recdate{July 29, 2022}

\abst{The axion search experiments based on haloscopes at the Center for Axion and Precision Physics Research (CAPP) of the Institute for Basic Science (IBS) in South Korea are performed in the frequency range from 1 GHz to 6 GHz. In order to perform the experiments in a strong magnetic field of 12 T and a large-volume cavity of close to 40 liters, we use He wet dilution refrigerators with immersed superconducting magnets. The measurements require continuous operation for months without interruptions for microwave component replacements. This is achieved by using different cryogenic engineering approaches including microwave RF-switching. The critical components, defining the scanning rate and the sensitivity of the setup, are the Josephson parametric amplifiers (JPA) and cryogenic low noise amplifiers (cLNA) based on high-electron-mobility-transistor (HEMT) technology. It is desirable for both devices to have a wide frequency range and low noise close to the quantum limit for the JPA. In this paper, we show a recent design of a 4-channel measurement setup for JPA and HEMT measurements. The setup is based on a 4-channel wideband noise source (NS) and is used for both JPA and HEMT gain and noise measurements. The setup is placed at 20 mK inside the dry dilution refrigerator. The NS is thermally decoupled from the environment using plastic spacers, superconducting wires and superconducting coaxial cables. We show the gain and noise temperature curves measured for 4 HEMT amplifiers and 2 JPAs in one cool-down.}

\kword{cryogenic noise source, superconducting circuit readout, quantum limited amplifier, Josephson parametric amplifier, microwave amplifier}

\begin{document}
\maketitle
\section{Introduction}
The axion search experiments based on haloscopes at the CAPP are performed at frequency ranges from 1 to 6 GHz in dilution refrigerators (DR) at high magnetic fields.
The key elements of the measurement setup are the microwave cavities\cite{Bru,Lee20,Jeong20,Kwon21,JinSu22}, JPAs\cite{Kutlu21} and HEMTs.
Along with the parameters of the cavity quality factor, volume, and the magnitude of the magnetic field, which determine the axion scanning speed, an important parameter is the sensitivity of the microwave chain. 
It is mainly defined by the intrinsic noise of cLNA.
Since the axion scanning requires operation over a wide frequency range, it is desirable to use wideband cLNAs.
The HEMT-based cLNAs are well known devices with high gain stability and a wide frequency range\cite{Wad09,Iv20}.
The noise temperature of cLNAs based on the HEMT technology is in the 1--5-K range at the ambient temperature of 4 K.

On the other side, in order to perform experiments with a noise level close to the quantum limit, we use the flux-driven JPAs\cite{Yamomoto08,Kutlu21}.
The main disadvantage of these is the narrow operation frequency range compared to broadband axion search experiments.
One solution for broadband axion search experiments near the quantum noise limit is using several JPAs in the measurement setup.
However, to carry out such an experiment, a preliminary experimental study of the operating parameters of a set of Josephson parametric amplifiers is required.
This study must be performed for individual JPA and HEMT measuring chains, including separate noise sources for each measurement channel. 
In this regard, it is important to design a multichannel variable noise source for LNA measurements, which operates at cryogenic temperatures over a wide microwave frequency range.
\begin{figure}
	\includegraphics[scale=0.5]{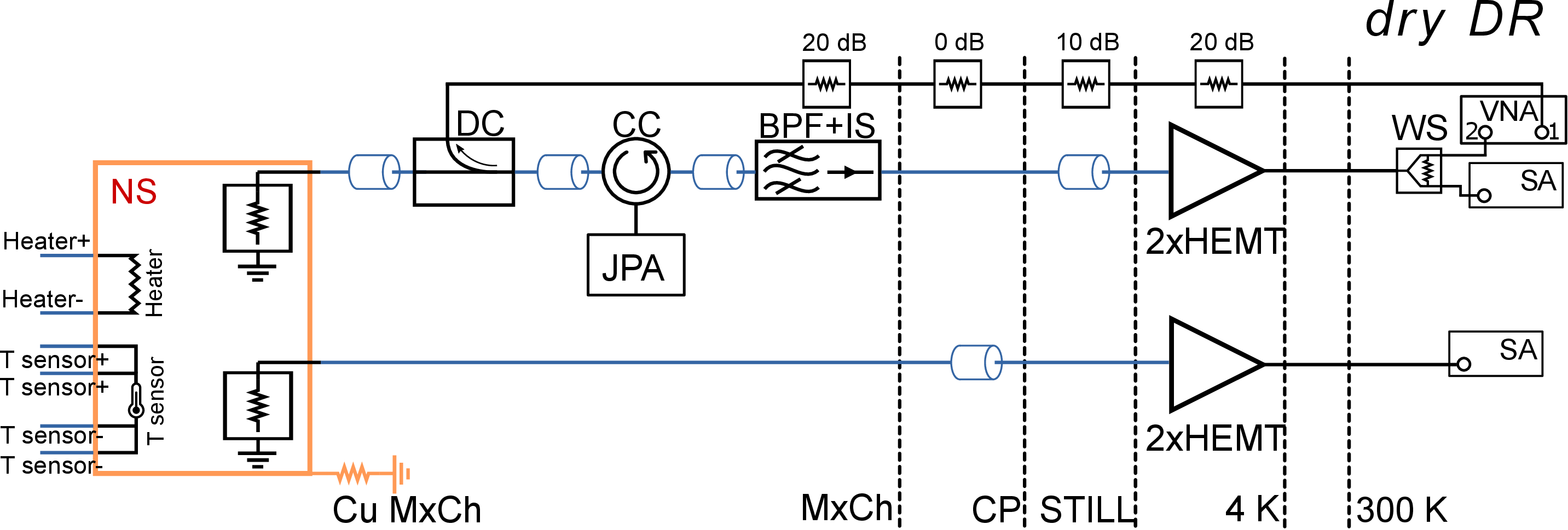}
	\centering
	\caption{Simplified 2-channel schematic of the JPA and HEMT measurement chains based on a 4-channel NS, DC: directional coupler, CC: cryogenic circulator, BPF+IS: band-pass filter and microwave isolator, WS: wideband microwave splitter, VNA: vector network analyzer, SA: spectrum analyzer, DR: dilution refrigerator, MxCh: mixing-chamber plate, and CP: cold plate. Blue color lines correspond to superconducting NbTi cables. The two other JPA and HEMT measurement channels were identical to the shown circuit.}
	\label{fig1}
\end{figure}

In this paper, we describe the four-channel cryogenic noise source dedicated to operate at temperatures from 10 mK up to 1 K, the NS design, and the main operating parameters.
The NS is installed at the mixing chamber stage of a dry DR in the JPA and HEMT measurement setup (see Fig.~\ref{fig1}).
Based on the designed NS, we performed an experimental study of the gain and noise temperature ($T_{N}$) for four HEMT-based cryogenic amplifiers and two JPAs at different frequency ranges. 
\section{Noise source}
\subsection{Design}
\begin{figure}
\includegraphics[scale=0.5]{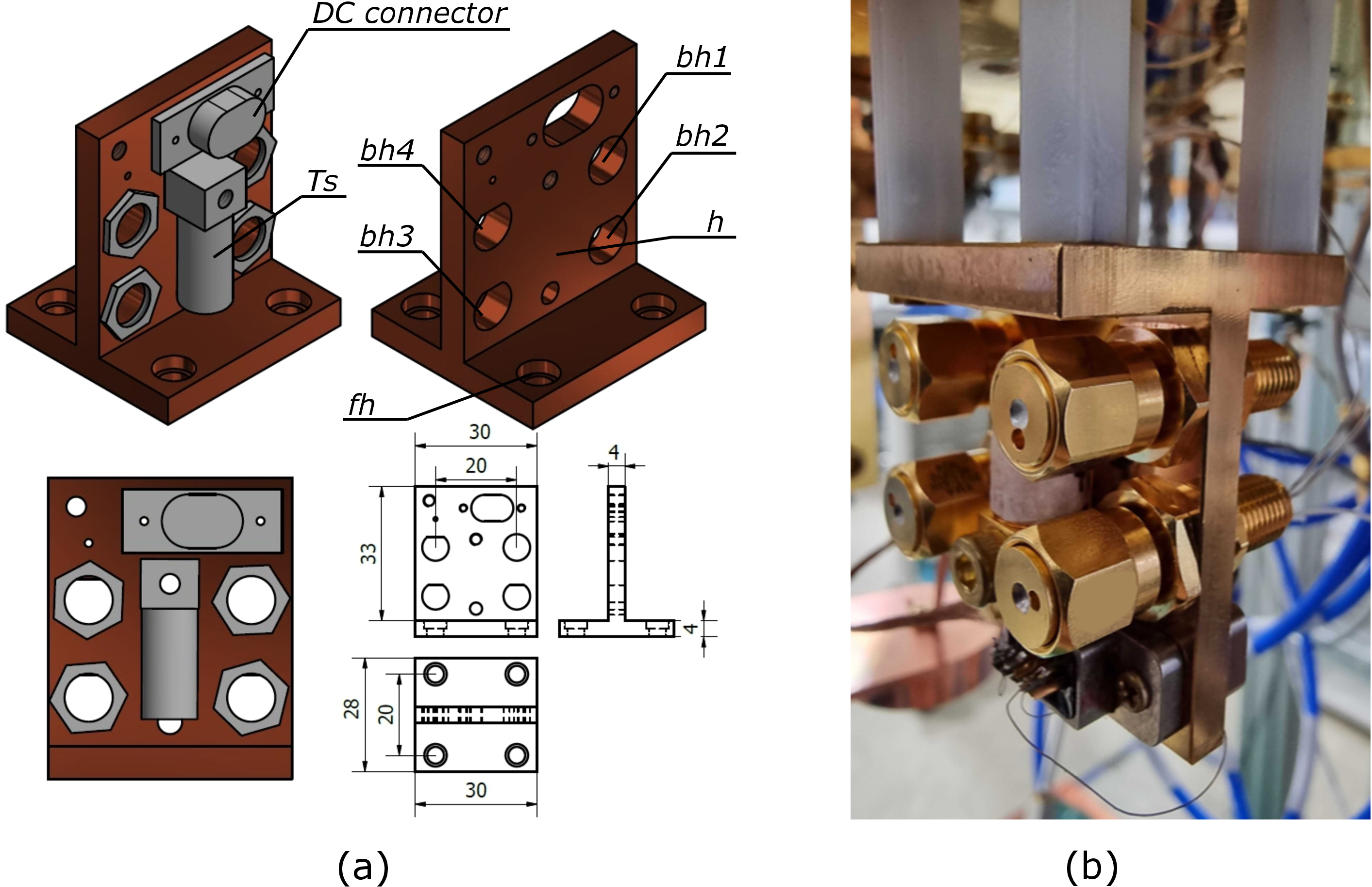}
\centering
\caption{The Noise source. (a) A technical drawing. The \textit{fh} and \textit{bh1--bh4}: are the feed through holes for fixing the NS to the mixing chamber plate and the holes for placing the SMA bulkheads, \textit{h}: the position of the heater, and \textit{Ts}: temperature sensor. (b) The view of the wideband noise source installed at the mixing chamber plate of the DR.}
\label{fig2}
\end{figure}

The noise source consists of wideband matched loads, a heater, a thermometer and connectors. 
All components were assembled on a solid copper holder, and the final construction is shown in Fig. \ref{fig2}a. 
The copper plates were coated with a 6 micron gold layer for better thermal anchoring of the Subminiature Version A (SMA) bulkhead adapters, D-Sub micro-D DC connector, thermal anchoring Cu wire, temperature sensor and heater to the NS. 
The noise source was designed to be installed at the mixing chamber plate of a BlueFors dry DR. 
Four feed-through holes for fastening the NS to the mixing chamber plate are shown as \textit{fh} in Fig. \ref{fig2}. 
The NS design is aimed to keep compact dimensions for minimizing the occupied space at the mixing chamber plate and provide sufficient heat capacity of the NS for physical temperature stabilization. 
The SMA bulkhead adapters were mounted into the feed-through holes of the NS, marked as \textit{bh1-bh4} in Fig. \ref{fig2}. 
The overall dimensions of the NS construction corresponds to 37 mm by 30 mm.  

Four bulkhead SMA adapters were used to provide a connection between the 50-Ohm cryogenic wideband matched load and the measurement chain for each channel. 
The wideband XMA cryogenic 50-Ohm terminations with maximum VSWR of 1.15 together with SMA bulkhead adapter provided 50-Ohm matched connections in the frequency range up to 18 GHz. 
The NS was connected to the input of the measurement chain using four 2.2-mm NbTi microwave coaxial cables.

Three superconducting NbTi twisted pair cables were used to connect the temperature sensor and the heater to the DC connector of the NS. 
They are marked as \textit{Ts}, \textit{h} and \textit{DC connector} in Fig. \ref{fig2}. 
The calibrated Ruthenium Oxide temperature sensor and heater were placed in the middle of the construction. 
The 372 AC resistance bridge and temperature controller of the Lake Shore Cryotronics was used for the precision 4-point temperature measurement. 
The heater of the NS was made using a 100-Ohm metal film resistor.
In order to avoid any heat conductance between the mixing chamber DC connector and the noise source DC connector, superconducting twisted pair cables were used.
The NS was fastened to the mixing chamber plate with four 20-mm long plastic spacers.
Since both DC and RF connections to the NS were made from superconducting cables, the only thermal anchoring of the NS to the mixing chamber plate was made by a copper wire weak link, shown as \textit{Cu} in Fig. \ref{fig1}.  
The picture of the wideband NS installed in the DR is shown in Fig.~\ref{fig2}b.

\subsection{Main parameters} 
The gain and noise temperature of two JPA + HEMT and two HEMT-based microwave chains were measured using the setup described above and shown in Fig.~\ref{fig1}.
They made it possible to measure the gain and noise temperature of 4 different microwave chains.
The two HEMT setups had the same configuration.
The JPA measurement setups had different microwave components according to the operating frequency range (see the upper chain in Fig.\ref{fig1}).
Each JPA chain contained one cryogenic circulator with frequency ranges corresponding to 0.9-1.25 GHz and 2.15-2.4 GHz.
Using the JPA setups, we conducted gain and noise temperature studies for two HEMTs and two JPAs. 
The other two chains were dedicated for HEMT measurements only.
All the measurements were performed during a single cool-down of the DR.

The Y-factor method with different noise source temperatures was used to determine the noise temperature of the HEMTs and JPAs.
The JPA noise measurements required additional characterization with the JPA \emph{ON} and \emph{OFF} modes at a fixed temperature of the NS\cite{Kutlu21}.
The \emph{ON} and \emph{OFF} modes could be explained by the operation of the JPA in the presence and absence of the pump signal. 
For the \emph{ON} mode, the JPA pump was on, resulting in the JPA acting as a linear amplifier. 
For the \emph{OFF} mode, the JPA pump was off, and the JPA reflected the input signal without amplification.

The physical temperatures of the NS were set to the values of $T_1= 100$~mK, $T_2 = 200$~mK, $T_3 = 300$~mK and $T_4 = 400$~mK for the each measurement chains shown in Fig.\ref{fig1}. 

\begin{figure}
\includegraphics[scale=0.33]{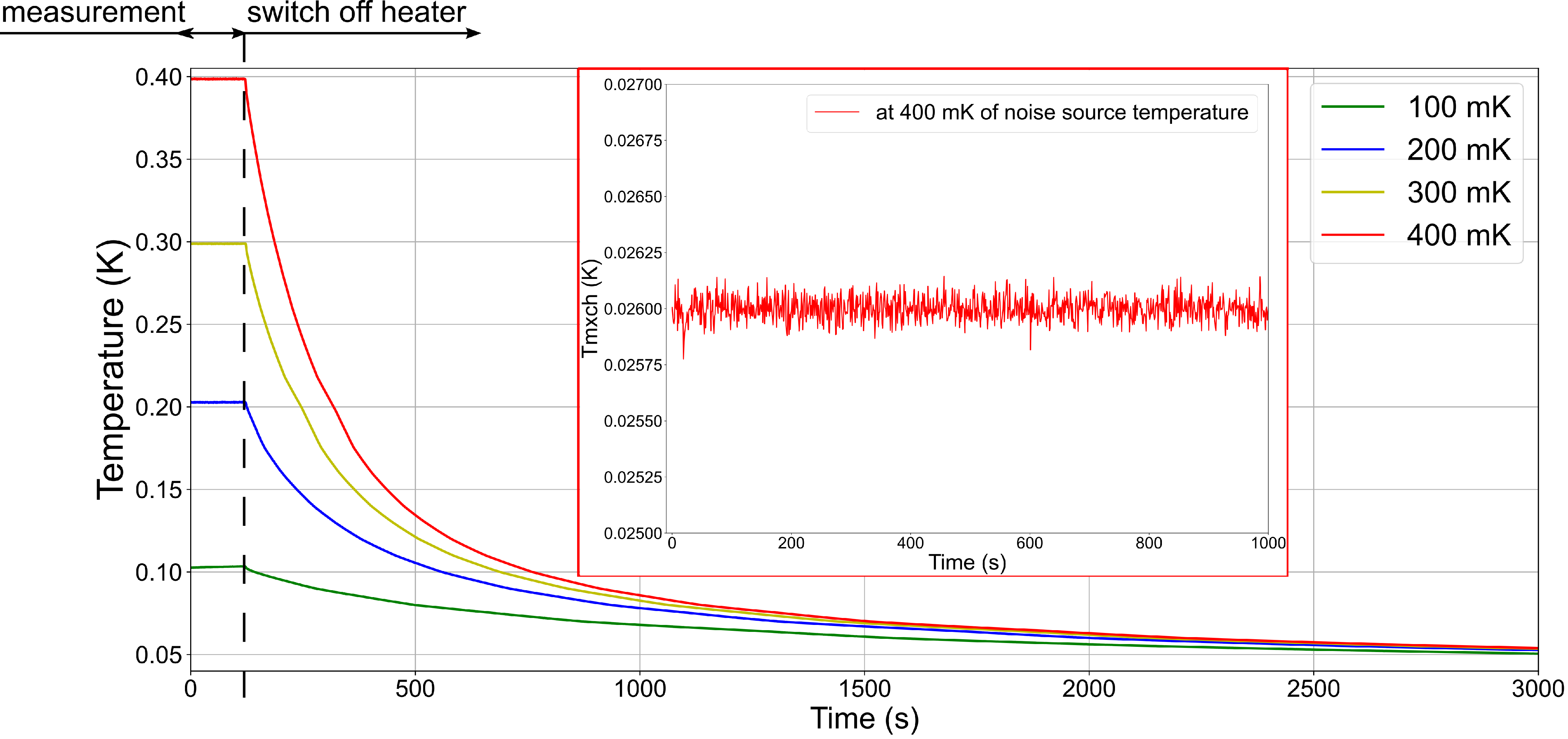}
\centering
\caption{NS temperature as a function of time for four temperatures from $T_1$ to $T_4$. The curves were placed with the same time offset. The vertical dashed line at 120 seconds shows where the heater was switched off. The inserted plot shows the mixing chamber temperature (Tmxch) depending on time at initial noise source temperature of 400~mK.}
\label{fig3}
\end{figure}

The base temperature of the mixing chamber plate and the NS were stabilized and corresponded to 18 mK at the beginning of the experiment.
The 372 AC PID temperature controller was used to set and stabilize the temperature of the NS during each measurement.
The time for the temperature stabilization at every noise measurement experiment after switching from $T_{i-1}$ to $T_i$ was 30 s.
After this time the temperature was observed to be stable.
Next, the measurement of the noise power spectrum was performed using a spectrum analyzer.
The measurement period was 2 minutes for each temperature point $T_i$.
The NS was cooled down to 50 mK temperature after the measurement at $T_4$.
A single measurement cycle of the power spectrum for NS temperatures from 100 mK to 400 mK with 100 mK steps, including the temperature settling time, the spectrum measurement time and the cooling down from 400 mK back to 100 mK time, took 23 minutes.
The temperature of the NS during and after each independent measurement at $T_i$ as a function of time is shown in Fig.~\ref{fig3}.
The temperature of the mixing chamber thermometer depending on time at initial noise source temperature of 400~mK is shown in the inserted plot of Fig.~\ref{fig3}.    
\subsection{NS application in LNA's measurements}
\begin{figure}
\includegraphics[scale = 1]{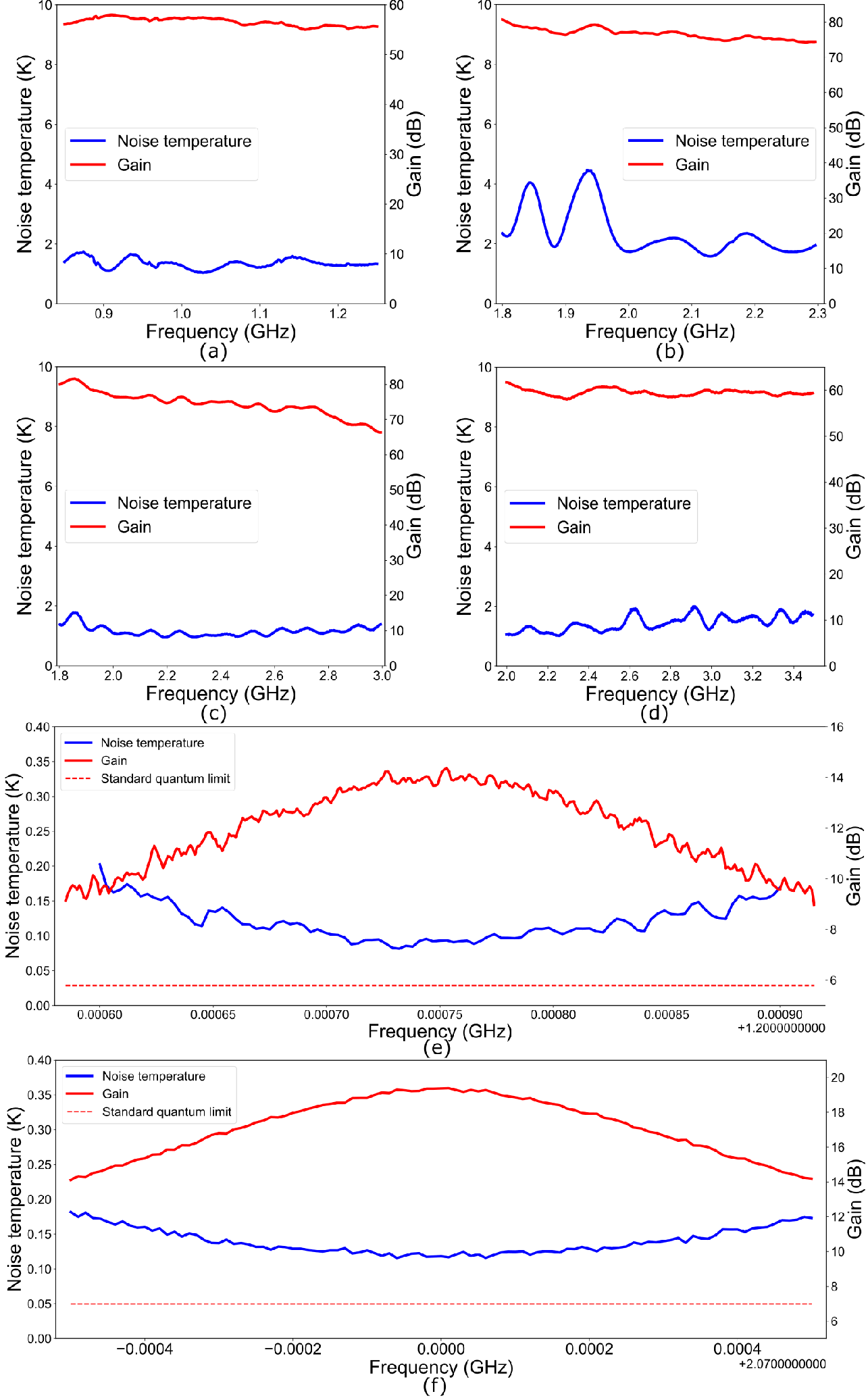}
\centering
\caption{The gain and $T\textsubscript{N}$ dependence on frequency for four HEMT-based cLNAs (a)-(d) and two JPAs~(e)-(f). The SQL is shown by dashed line.}
\label{Fig4}
\end{figure}

Four measurement chains had commercially available HEMT cLNAs.
In order to obtain a gain higher than 60 dB for each measurement chain two HEMT cLNAs were installed.
The 8 HEMT cLNA assembly was placed at the 4K stage of the dry DR.
The temperature of the assembly was stabilized at 3.6 K.
The Y-factor method with 4 different temperatures was used for the experimental study of the gain and noise properties for 4 HEMT cLNAs based on the setup shown in Fig.~\ref{fig1}.
The gain and noise temperature of these chains were measured in the JPA \emph{OFF} mode, the results are shown in Figs.~\ref{Fig4}(a) and~\ref{Fig4}(b).
The gain and $T_n$ for the HEMT measurement chains 3 and 4 are shown in Figs.~\ref{Fig4}(c) and~\ref{Fig4}(d).
We used different second-stage cryogenic amplifiers which resulted in different total gain values.  
The wideband superconducting NbTi coaxial cables were used for reducing losses in the measurement chains.
They are marked with the blue color in Fig. \ref{fig1}.

The Josephson parametric amplifiers were fabricated at the RIKEN Center for Quantum Computing (RQC).
The full JPA noise measurement procedure required additional measurements. 
First, the base line was measured in JPA \emph{OFF} mode using an additional microwave channel with high attenuation and a cryogenic directional coupler.
The total losses of the line corresponded to 70 dB at cryogenic temperatures for both JPA chains (see Fig.~\ref{fig1}).
Next, the gain curves of both JPAs in the JPA \emph{ON} mode were measured.
The noise temperature was estimated by comparing the power spectra in the JPA \emph{OFF} and \emph{ON} modes.
The measurement results are shown in Fig.~\ref{Fig4}(e) and~\ref{Fig4}(f).
The minimum noise temperatures were 80~mK and 130~mK with maximum gain values of 14~dB and 19~dB for chain one and chain two respectively. 
The standard quantum limit (SQL) for the linear phase-insensitive amplifiers given by \(T=hf/2k_B\) \cite{Caves82} is shown in Fig.~\ref{Fig4}(e) and~\ref{Fig4}(f).
Where, \textit{h} is Planck's constant, \textit{f} is the amplifier's input operating frequency, \textit{$k_B$} is the Boltzmann's constant.
\section{Conclusion}
A four-channel wideband noise source was designed and used for gain and noise-temperature measurements of HEMT-based cLNAs and JPAs. 
The operating temperature of the noise source ranges from 20~mK to 1~K.
Based on this noise source, we performed gain and noise-temperature measurements of two Josephson parametric amplifiers at different frequencies. 
The noise temperature of the JPAs was in a range from 80 mK to 130 mK at 1.2~GHz and 2~GHz respectively. 
This noise source is suitable for gain and noise-temperature measurements for several cLNAs including Josephson parametric amplifiers with noise close to the quantum limit in a frequency range up to 18~GHz. 
\section*{Acknowledgement}
This work is supported in part by the Institute for Basic Science (IBS-R017-D1) and JST ERATO (Grant No. JPMJER1601). Arjan F. van Loo was supported by the JSPS postdoctoral fellowship.

\end{document}